\documentclass{aa}  

\usepackage{graphicx}
\usepackage{tabularx}
\usepackage{xcolor}
\usepackage{keyval}

\usepackage{txfonts}

\usepackage{hyperref}

\begin{document}

   \title{Updated predictions for gravitational wave emission from TDEs for next generation observatories}

   \author{M. Toscani
          \inst{1,2 \thanks{martina.toscani@unimib.it}}
          \and
          L. Broggi
          \inst{1,2}
          \and
          A. Sesana
          \inst{1,2}
          \and
          E. M. Rossi
          \inst{3}
}
    \institute{Dipartimento di Fisica G. Occhialini, Università degli Studi di Milano-Bicocca, piazza della Scienza 3, Milano, Italy
    \and
    INFN, Sezione di Milano-Bicocca, Piazza della Scienza 3, I-20126 Milano, Italy
    \and Leiden Observatory, Leiden University, PO Box 9513, NL-2300 RA Leiden, the Netherlands
             }


 \abstract
{In this paper, we investigate the gravitational wave (GW) emission from stars tidally disrupted by black holes (TDEs), using a semi-analytical approach. Contrary to previous works where this signal is modeled as a monochromatic burst, we here take into account all its harmonic components. On top of this, we also extend the analysis to a population of repeated-partial TDEs, where the star undergoes multiple passages around the black hole before complete disruption. For both populations, we estimate the rate of individual GW-detections considering future observatories like LISA and a potential deci-Hertz (dHz) mission, and derive the GW background from these sources. Our conclusions, even if more conservative, are consistent with previous results presented in literature. In fact, full disruptions of stars will not be seen by LISA but will be important targets for dHz observatories. In contrast, GWs from repeated disruptions will not be detectable in the near future.}


\maketitle

\section{Introduction}
In the classical scenario, a tidal disruption event (TDE; see, e.g., \citealt{Rees:88aa,Phinney:89aa}; for a recent review \citealt{Rossi:21aa} and references therein) is an extreme phenomenon in which a star orbiting around a massive black hole (BH) is shred apart by its tides. The aftermath of the disruption is visible thanks to luminous (possibly super-Eddington) flares produced by accretion of the stellar debris onto the BH. Since the 1990s, we have detected these events through different wavelengths of the electromagnetic spectrum (see, e.g., recent reviews by \citealt{Vanvelzen:20aa, Saxton:21aa} and references therein) and this number is expected to increase significantly in the near future thanks to upcoming sensitive detectors \citep{Gomez:23aa}, like The Rubin Observatory Legacy Survey of Space and Time (LSST; \citealt{Lsst:19aa}) and the Ultraviolet Transient Astronomy Satellite \citep[ULTRASAT;][]{Ultrasat:24aa}. 

Lately, there has also been a growing interest towards a distinct class of TDEs, where the star survives the first passage at the pericenter and undergoes multiple orbits before complete disruption (see e.g., \citealt{Bortolas:23aa}). The interest in these sources is also motivated since they could be linked to recently observed quasi-periodic eruptions \citep[QPEs;][]{Itai:23aa}, that still remain without a widely accepted explanation \citep[see, e.g., ][for an alternative model]{Franchini:23aa}.

TDEs are also expected to produce gravitational waves (GWs) in the low frequency regime (mHz-dHz). Previous works \citep{Kobayashi:04aa, Toscani:20aa,Pfister:22aa,Toscani:22aa, Toscani:23aa}  have already explored these signals, generally concluding that while for LISA \citep{Colpi:24aa} - the first GW space-based detector set to launch in 2035 - it is unlikely to observe these events, the situation will significantly improve with potential future interferometers optimized to the dHz frequency band. 

In this paper, we refine the estimates for future space-based gravitational observatories of both TDE detection rates, as well as of the GW background (GWB) from these sources, building on previous works by \citet{Toscani:20aa} and \citet{Pfister:22aa}. First, we improve the strain estimate accounting for all harmonics contributing to the GW emission, following the formalism for parabolic encounters presented in \citet{Berry:10aa}. Second, we extend this study by incorporating a population of repeated partial disruptions, based on the model proposed in \citet{Broggi:24aa}, to make the analysis more comprehensive. This scenario could be particularly intriguing from a GW perspective, as repeated passages of the star around the BH allow for the accumulation of the signal-to-noise ratio (SNR), thus potentially improving its detectability compared to complete disruptions, which feature, by definition,  one single passage. 
For clarity, we introduce the following terminology: full TDEs (fTDEs), referring to classic stellar disruptions, and repeated partial TDEs (rpTDEs), describing the scenario where the star undergoes multiple orbits around the BH.

The structure of the paper is the following: in Sections \ref{sec:ftdes} and \ref{sec:rptdes} we illustrate the calculations for fTDEs and rpTDEs respectively, computing the strain of individual events and the overall GWB.  
In Section \ref{sec:discussion} we discuss our findings and compare them to previous studies, and finally in Section \ref{sec:conclusions} we draw our conclusions.

In this work, we consider a $\Lambda$CDM model with $H_0=70\,\text{km/s/Mpc}$ and $\Omega_{\rm m}=0.3$; we refer to the gravitational constant, the speed of light in vacuum, solar mass and radius as $G$, $c$, $\text{M}_{\odot}$ and $\text{R}_{\odot}$ respectively.

\section{Full disruptions}
\label{sec:ftdes}

Before diving into calculation, we introduce some key concepts of dynamics of dense stellar systems surrounding a massive BH. In particular, we introduce the loss cone and illustrate the difference between full and empty loss cone regimes, which helps connecting stellar dynamics with the nature of TDEs. 

The loss cone (\citealt{Frank:76aa}, recent review by \citealt{Stone:20aa}) is the region in the (specific) angular momentum and energy space where stars can pass close enough to the BH to be disrupted or captured. fTDEs occur in the full loss cone regime (also referred to as pinhole regime), where stars take large angular momentum steps over the orbital period and are scattered in and out of the loss cone within subsequent pericenter passages. This is the typical scenario of stars weakly bound to the BH, approaching onto almost radial orbits. If they are inside the loss cone at the pericenter, a full disruption happens. On the other hand, stars deeply bound to BH typically fall in the empty loss cone regime, where they need many orbital revolutions to change significantly their angular momentum, so that they graze the loss cone and their disruption is gradual. Hence, in this regime we have rpTDEs.

To be more quantitative, we can introduce the dimensionless parameter
\begin{align}
    q= \left(\frac{\Delta J}{J_{\rm lc}}\right)^2,
\end{align}
where $\Delta J$ is the step in the angular momentum and $J_{\rm lc}=(2GM_\bullet 
R_{\rm p})^{1/2}$ is the angular momentum at the surface of the loss cone, being $M_\bullet$ the BH mass and $R_{\rm p}$ the stellar pericenter. If $q>1$, we are in the pinhole regime, while for $q<1$ the empty loss cone regime holds.
In the rest of this Section, we analyze fTDEs.

\begin{figure}
    \centering
    \includegraphics[width = 0.5\textwidth]{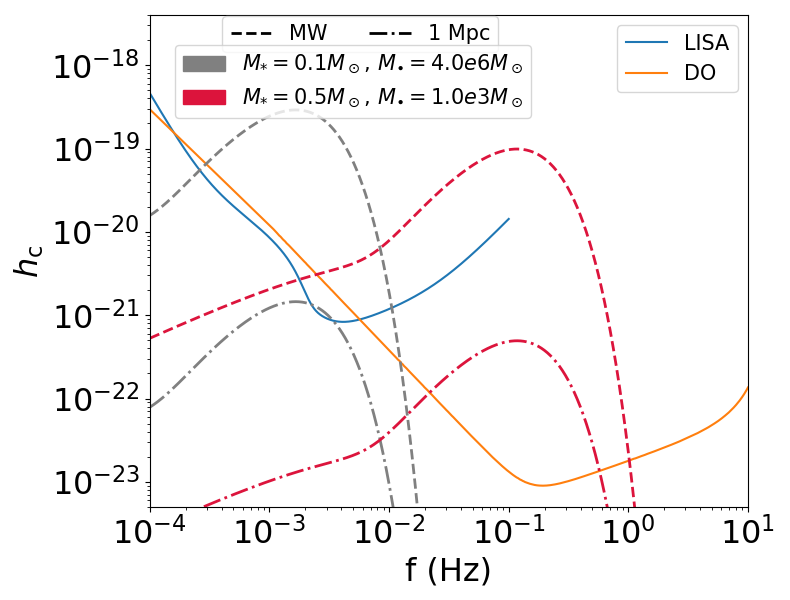}
    \caption{GW signal from a fTDE of a 0.1M$_\odot$ MS star disrupted by a $4\times 10^6\text{M}_{\odot}$ BH (grey lines) and GW signal from a fTDE of a WD of $0.5\text{M}_{\odot}$ disrupted by a $10^3\text{M}_{\odot}$ BH (red lines). We consider two possible source distances: 8 kpc (dashed lines) and 1 Mpc (dash-dot lines). In blue and orange we plot the noise amplitudes of LISA and DO, respectively.
    }
    \label{fig:signal_examples}
\end{figure}

\begin{figure}
\centering
\includegraphics[width = 0.5\textwidth]{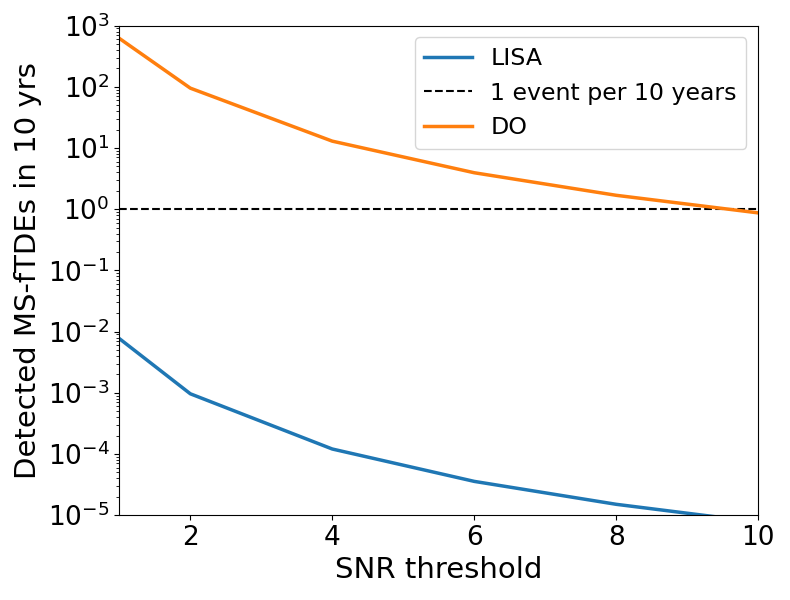}
    \caption{Detection rates of MS-fTDEs plotted for different SNR thresholds. The blue curve is calculated considering LISA as the detector, while the orange curve is obtained considering DO. The observation time for both instruments is assumed to be 10 years. The dashed black line is one event over 10 years.}
    \label{fig:MS_ftde}
\end{figure}
\begin{figure}
\centering
\includegraphics[width = 0.5\textwidth]{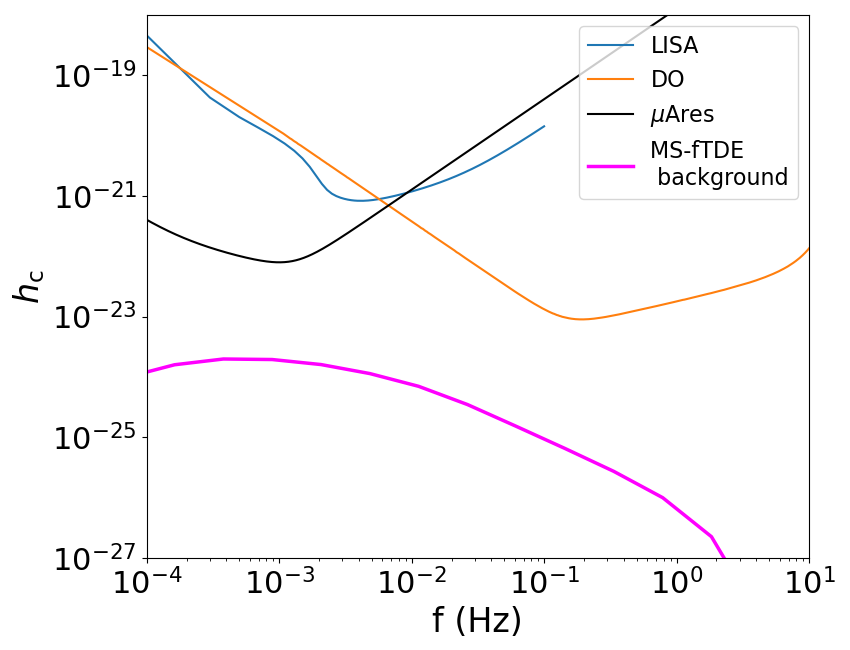}
    \caption{MS-fTDEs background (magenta curve) plotted with respect to the sensitivity curves of LISA (blue line), DO (orange line) and $\mu$Ares (black line).}
    \label{fig:back_ftde}
\end{figure}

\subsection{Two examples}
To get an order of magnitude idea of the GW signals we expect from these sources, we start by considering two examples\footnote{In this work, we consider the following mass-radius relation for MS stars: $R_*/R_{\odot}\approx M_*/M_\odot$. While for the WD, we consider a fixed mass and radius of $0.5\text{M}_{\odot}$ and $0.01\text{R}_{\odot}$.}: i) a fTDE of a MS star with mass $M_*=0.1\,\text{M}_{\odot}$, destroyed by a BH of size $M_\bullet =4\times 10^6\text{M}_{\odot}$ and ii) a fTDE of a WD with mass $M_*=0.5\,\text{M}_{\odot}$ and BH $M_\bullet=10^{3}\text{M}_{\odot}$. We examine two possible source distances: 8 kpc (the distance of the Galactic center of the Milky Way) and 1 Mpc (about the size of the Local Group of galaxies).

Considering that $|\tilde{h}(f)|$ is the Fourier transform of the GW strain, the inclination-polarization averaged characteristic strain reads 
\begin{align}
    h_{\rm c}^2=4|\tilde{h}(f)|^2f^2=\frac{G}{\pi^2 c^3 \chi^2}\frac{dE}{df},
    \label{eq:ft_strain}
\end{align}
where $\chi$ is the comoving distance \citep{Hogg:99aa}. For the GW energy spectrum, $dE/df$, we use the expression for parabolic encounters illustrated in \cite{Berry:10aa}
\begin{align}
 \frac{dE}{df}=\frac{4\pi^2G^3}{5c^5}\frac{M_{\bullet}^2M_*^2}{R_{\rm p}^2}\ell\left(\frac{f}{f_{\rm c}}\right), 
 \label{eq:energy_spectrum}
\end{align}
with $\ell(f/f_{\rm c})$ containing the contributions from all the harmonics and $f_{\rm c}$ being the Keplerian frequency of the circular orbit with radius equal to the stellar pericenter, $R_{\rm p}$.

The signal-to-noise ratio, SNR, for a GW is given by \citep[]{Maggiore:08aa}
\begin{align}
    \text{SNR}^2 = 4\int_0^{\infty}df \frac{|\tilde{h}(f)|^2}{S_{n}(f)},
    \label{eq:snr}
\end{align}
where $S_{n}(f)$ is the sky-averaged power spectral density (PSD) of the instrument. In this work, we consider two reference detectors: i) LISA, assuming a mission duration of 10 years (\citealt{Colpi:24aa}), and ii) an optimal model for the dHz observatory (DO; in particular we refer to the instrument labeled as DO-optimal in \citealt{Sedda:20aa}). 

In Fig. \ref{fig:signal_examples}, we plot $h_{\rm c}$ with respect to the noise amplitudes $(\sqrt{fS_{\rm n}})$ of both instruments, and we show that this signal is potentially detectable. In fact, the MS star emission could be seen by LISA (SNR $\sim 193$) and also by DO (SNR $\sim 91$) within our Galaxy\footnote{In this study, we define detectable TDEs those with a SNR $\geq 1$ (for consistency with \citealt{Pfister:22aa}), basically treating in the same way sure and marginal detections. We acknowledge that this threshold is too low for practical detection. However, since our goal is not to carry out a data analysis study but to provide a guideline for future research, we decided to follow this approach.}, however, it becomes undetectable to both instruments once the source is located outside the Milky Way. Instead for the WD, since the signal peaks around 0.1 Hz where DO is particularly sensitive, it can be seen by this instrument with a relatively high SNR ($\sim 47$) even at a distance of 1 Mpc. This is not the case for LISA, whose sensitivity is insufficient to detect the signal beyond a few kpc.

These estimates alone do not really inform us about the detectability of these sources. To understand whether they can be observed, we need to take into account the effective number of TDEs occurring in the Universe, as we illustrate in the following paragraphs.

\subsection{Main sequence stars: individual detection rates}
\label{sec:ftdes_ms}
The rate of detected MS-fTDEs per bin of redshift $z$, BH mass $M_\bullet$, stellar mass $M_*$ and pericenter $R_{\rm p}$ can be estimated as \citep{Toscani:20aa, Pfister:22aa}
\begin{align}
    \dot 
N^{\rm fTDEs}_{\rm det}= &\int dz \int dM_{\bullet}\int dM_{*} \int d R_{\rm p} \nonumber \\
&\frac{d^2\Gamma(M_{\bullet})}{dM_* dR_{\rm p}}\times \frac{4\pi c\chi^2(z)}{H(z)}\times\frac{\Phi(M_{\bullet},z)}{1+z}\times \Theta(z,M_{\bullet},M_{*},R_{\rm p}).
\label{eq:ms_ftde_rate}
\end{align}

The first term on the RHS of Eq.~\eqref{eq:ms_ftde_rate} is the differential TDE rate per galaxy, that we express as

\begin{align}
  \frac{d^2\Gamma(M_{\bullet})}{dM_* dR_{\rm p}}= \Gamma(M_{\bullet})\mathcal{F}(M_\bullet)\phi(M_*)\psi(R_{\rm p}), 
\label{eq:tde_diff_rate}
\end{align}
where $\Gamma(M_{\bullet})$ is the TDE rate as a function of the BH mass, that we take from the recent work by \citet{Chang:24aa}
\begin{align}
  \Gamma(M_\bullet)=\begin{cases}
      1.2\times 10^{-4}\text{yr}^{-1}\left(\frac{M_\bullet}{10^6 \text{M}_{\odot}}\right)^{0.9}\,\,\,\,M_\bullet \leq 10^6 \text{M}_{\odot},\\
        1.2\times 10^{-4}\text{yr}^{-1}\left(\frac{M_\bullet}{10^6 \text{M}_{\odot}}\right)^{-0.9} \,\,\,\,M_\bullet > 10^6 \text{M}_{\odot},
  \end{cases}  
\end{align}
$\mathcal{F}(M_{\bullet})$ is the fraction of TDEs occurring in the pinhole regime \citep[Eq. 29]{Stone:16aa}, $\psi(R_{\rm p})$ is the uniform pericenter distribution (in agreement with loss cone theory) and $\phi(M_*)$ is the Kroupa initial stellar mass function \citep{kroupaVariationInitialMass2001} 
\begin{align}
\phi(M_*)=\mathcal{N}
\begin{cases}
    {M_*}^{-0.3}\,\,\,\,\,\,\,\,\,\,\,\,\,\,\,\,\,\,\,\,\,\,\,\,\,\,\,\,\,\,\,\,\,\,\,\, M_* \leq 0.08 \text{M}_{\odot},\\
      {M_*}^{-1.3} \,\,\,\,\,\,\,\,\,\,\,\,\,\,\,\,\,\,\,\,\,\,\, 0.08\text{M}_{\odot}< M_* \leq 0.5\text{M}_{\odot},\\
      {M_*}^{-2.3} \,\,\,\,\,\,\,\,\,\,\,\,\,\,\,\,\,\,\,\,\,\,\,\,\,\, 0.5 \text{M}_{\odot} < M_* \leq 1\text{M}_{\odot},\\
      {M_*}^{-2.7} \,\,\,\,\,\,\,\,\,\,\,\,\,\,\,\,\,\,\,\,\,\,\,\,\,\,\,\,\,\,\,\,\,\,\,M_* >1\text{M}_\odot ,
    \label{eq:stellar_mass_function}
\end{cases}
\end{align}
where the constant $\mathcal{N}$ is chosen to have a continuous function, normalized on the interval $(0,\infty)$. $\Phi(M_{\bullet},z)$ is the BH mass function that, following \citet{Pfister:22aa}, 
we model as a redshift-dependent double Schechter function. The function parameters are obtained by fitting the galaxy mass function 
of the COSMOS field (Eq. 4 and Table I from \citealt{Davidzon:17aa}) and then by converting the galaxy mass into the central BH mass using the relation by \citet{Reines:15aa}. Finally, $\Theta$ is a step function, being 1 if the gravitational signal from the TDE  with $z$, $M_\bullet$, $M_*$, $R_{\rm p}$ is above threshold, 0 otherwise.

In this calculation, we consider the following ranges for the variables:
\begin{itemize}
    \item $0 \lesssim  z \leq 3$;
    \item $10^{4}\text{M}_{\odot} \leq M_{\bullet} \leq 10^{7}\text{M}_{\odot}$;
    \item $0.1\text{M}_{\odot} \leq M_* \leq 1\text{M}_{\odot}$ 
    \item $R_{\rm sch}(M_\bullet)\leq R_{\rm p} \leq R_{\rm t}(M_\bullet, M_*)$, with $R_{\rm sch}$ Schwarzschild radius of the BH and $R_{\rm t}$ tidal radius.
\end{itemize}

The result is illustrated in Fig.~\ref{fig:MS_ftde}, where we plot the detection rates as functions of the SNR thresholds for LISA (blue curve) and DO (orange curve). The dashed horizontal black like is one detection in 10 years. We see that while for LISA no detections of these events are expected, the situation is much better for DO. In this case, these sources could reach signal-to-noise ratios of up to $\sim 10$, with the number of detectable events varying from tens to hundreds, depending sensitively on the detection criteria adopted. 

\subsection{Main sequence stars: the background}
The GW signal from the entire population of MS-fTDEs can be expressed as \citep{Phinney:01aa, Toscani:20aa}
\begin{align}
    h_{\rm c, pop}^2=\frac{G}{c^3\pi^2}\times\frac{1}{f}\times \int_{0}^{\infty}dz \frac{1}{\chi^2}\frac{d\dot{N}^{\rm tde}}{dz} \left(\frac{dE}{df}\right),
    \label{eq:MS-background}
\end{align}
with
\begin{align}
    \frac{d\dot{N}^{\rm tde}}{dz}\frac{dE}{df}=&\int dM_{\bullet}\int dM_* \int dR_{\rm p} \frac{d^{4}\dot 
N^{\rm fTDEs}}{dzdM_{\bullet}dM_{*}dR_{\rm p}} \nonumber  \\ 
& \times \frac{4\pi^2G^3}{5c^5}\frac{M_{\bullet}^2M_*^2}{R_{\rm p}^2}\ell\left(\frac{f}{f_{\rm c}}\right),
\end{align}
where the differential rate is the same as the integrand in Eq.~\eqref{eq:ms_ftde_rate} without the $\Theta$ term, since now we also want to include the sources below the threshold of detectability. 

The result is illustrated in Fig. \ref{fig:back_ftde}, where the magenta curve represents the background. Compared to the previous plots, here we also include a third observatory,  $\mu$Ares \citep{muares:21aa}, mainly as a reference given that this part of the frequency window (0.0001-0.001 Hz) will be heavily dominated by backgrounds from different astrophysical sources (Perego et al. in prep.). 
The plot shows that the signal lies orders of magnitude below the sensitivity curves of the three considered instruments. Nevertheless, the most informative way to assess its detectability is by computing its SNR, \textcolor{black}{accounting for the observation time of the instrument. Following Eq. (7) of \citet{Sesana:16aa}, assuming $T=10$ years and $\gamma=1$, we get an SNR smaller than 1 in all cases, meaning that the signal will remain undetectable for the future space-based detectors considered in this study.}

Finally, it is also worth noting that the signal is much lower than expected in \cite{Toscani:20aa}. However, a direct comparison is not straightforward due to differences in the integration limits adopted in the two calculations. To address this, in Sec. \ref{sec:discussion}, we carry out a detailed analysis to actually understand the origin of the difference between the two estimates.

\subsection{White dwarfs: individual detection rates}
\label{sec:ftdes_wd}
We now focus on the case of WD-fTDEs. This scenario is intriguing since, as already pointed out in \citealt{Toscani:20aa}, due to their compactness they generate gravitational signals comparable in magnitude to those produced by MS stars (cf. Fig. \ref{fig:signal_examples}), although the BHs involved in the disruption are smaller.\footnote{This is consistent with the traditional description of a TDE, considered in primis as an electromagnetic source, thus requiring the stellar pericenter to be outside the BH horizon to be able to detect electromagnetic emission.}

Here in particular we distinguish between two classes of events, involving WDs disrupted by  different populations of BHs: i) intermediate mass BHs in the center of dwarf galaxies and ii) intermediate mass BHs residing in globular clusters (GCs). For case i) the rate of detected WD-fTDEs is the same as in Eq.~\eqref{eq:ms_ftde_rate} apart from the differential TDE rate per galaxy, which reads
\begin{align}
\frac{d^2\Gamma(M_{\bullet})}{dM_* dR_{\rm p}}= \Gamma(M_{\bullet})\mathcal{F}(M_\bullet)\delta(M_*)\psi(R_{\rm p}).
\label{eq:diff_tde_dwarf}
\end{align}
The stellar mass distribution is replaced with a $\delta$ function since we consider a WD population monochromatic in mass (with $M_*=0.5\text{M}_{\odot}$, $R_*=0.01\text{R}_{\odot}$). In this case, we assume BHs within the range $10^{3}\text{M}_{\odot} \leq M_\bullet \leq 10^{5}\text{M}_{\odot}$, while for $z$ and $R_{\rm p}$ the limits are defined as in Sec. \ref{sec:ftdes_ms}.

For case ii), the differential TDE rate per galaxy becomes
\begin{align}
\frac{d^2\Gamma(M_{\bullet})}{dM_* dR_{\rm p}}= \Gamma(m_{\bullet})\mathcal{F}(m_\bullet)N_{\rm gc}(M_\bullet)\delta(M_*)\psi(R_{\rm p}).
\end{align}
We have indicated the BH within the GC as lower case $m_\bullet$, to easily distinguish it from the one within the galactic core. $N_{\rm gc}(M_\bullet)$ is the number of GCs as a function of the central galaxy BH \citep{Harris:11aa}
\begin{align}
    N_{\rm gc}= \left( \frac{M_\bullet}{4.07\times 10^5\text{M}_{\odot}}\right).
\end{align}
This formula requires that $M_\bullet \gtrsim 10^{6}\text{M}_{\odot}$ in order to have at least one GC per galaxy; in particular we consider the range $10^{6}\text{M}_{\odot}\leq M_\bullet \leq 10^8\text{M}_{\odot}$. We assume that the occupation fraction of intermediate mass BHs in GCs is 1 and we take a fixed size of $m_\bullet=10^3\text{M}_{\odot}$. The limits for $z$ and $R_{\rm p}$ are defined as in Sec.~\ref{sec:ftdes_ms}.

\begin{figure*}
    \centering
    \begin{minipage}{0.49\textwidth}
        \centering
        \includegraphics[width=\linewidth]{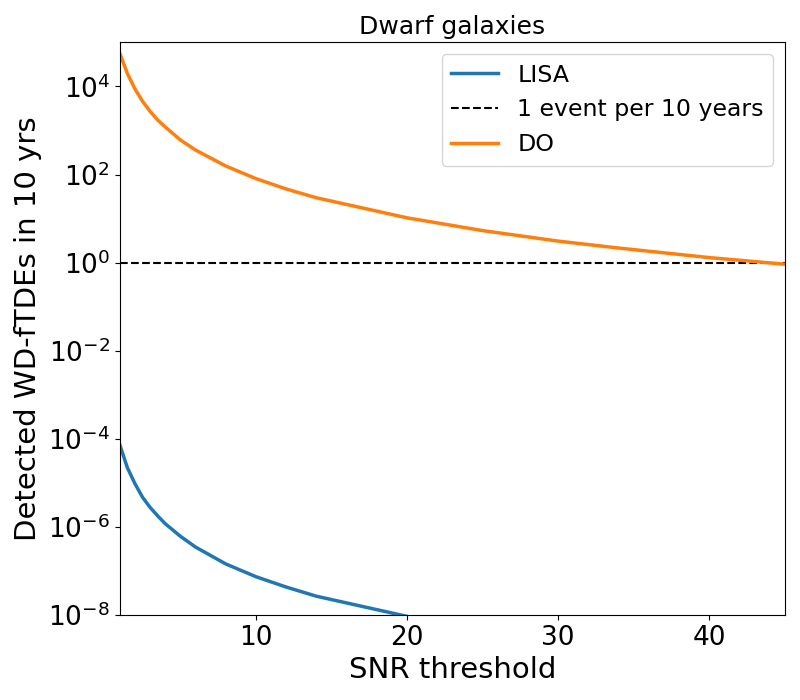}
    \end{minipage}
    \begin{minipage}{0.49\textwidth}
        \centering
        \includegraphics[width=\linewidth]{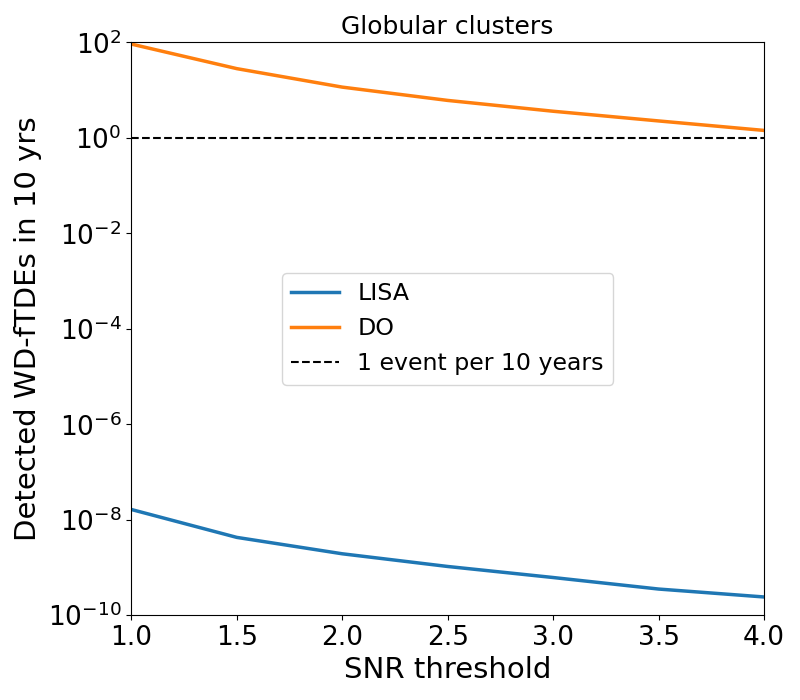}
    \end{minipage}
    \caption{Detection rates of WD-fTDEs from dwarf galaxies (left panel) and from globular clusters (right panel) plotted for different SNR thresholds. The blue curve is calculated considering LISA as the detector, while the
    orange curve is obtained considering DO. The observation time for both instruments is assumed to be 10 years. The dashed black line is one event over 10 years. \textcolor{black}{The occupation fraction is 1 for both scenarios.}}
    \label{fig:wd_individual_det}
\end{figure*}

The results are illustrated in Fig.~\ref{fig:wd_individual_det}, on the left panel for WD-fTDEs in dwarf galaxies and on the right panel for WD-fTDEs from GCs. While LISA will not be able to see any emission from these sources, DO may marginally detect up to a few tens of fTDEs from GCs with SNR between 1 and 4. As for fTDEs from dwarf galaxies, the detection prospects improve significantly (with maximum SNR $\sim 45$), with expected observations ranging from tens to hundreds depending on the SNR threshold.

\subsection{White dwarfs: the background}
The background from WD-fTDEs is expressed with the same formula as in Eq.~\eqref{eq:MS-background}, yet this time we write the rate of sources per redshift bin considering the differential rates described in Sec.~\ref{sec:ftdes_wd}.

\begin{figure*}
    \centering
    \begin{minipage}{0.49\textwidth}
        \centering       \includegraphics[width=\linewidth]{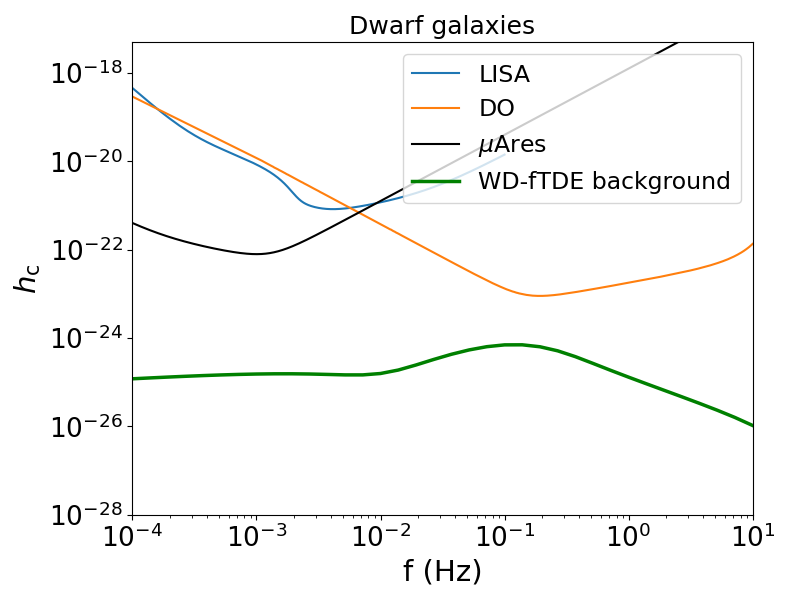}
    \end{minipage}
    \hfill
    \begin{minipage}{0.49\textwidth}
        \centering
        \includegraphics[width=\linewidth]{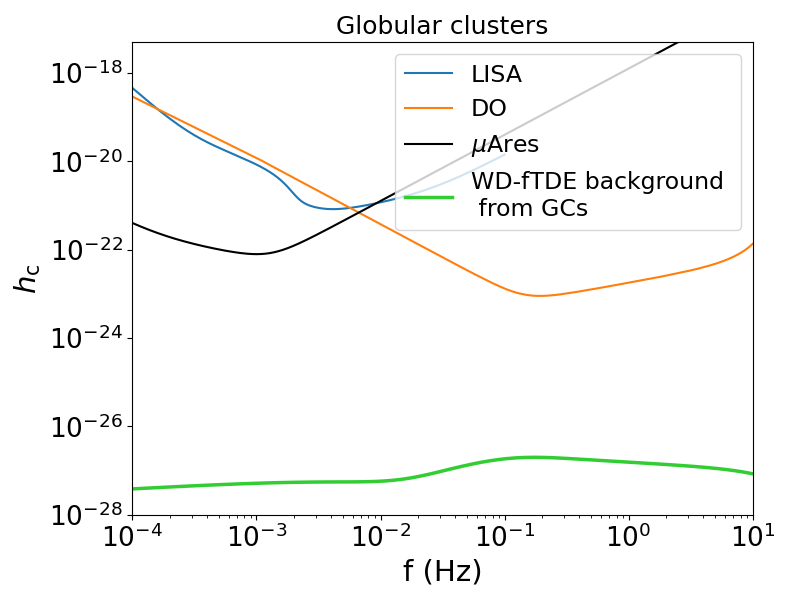}
    \end{minipage}
    \caption{WD-fTDEs background from dwarf galaxies (dark green line, left panel) and WD-fTDEs background from globular clusters (light green line, right panel). We assume that the mass of the intermediate mass BH in each globular cluster is $10^3\text{M}_{\odot}$ and \textcolor{black}{that the occupation fraction is equal to 1 for both scenarios}. Both signals are plotted with respect to the sensitivity curves of LISA (blue line), DO (orange line) and $\mu$Ares (black line).}
    \label{fig:wd_back}
\end{figure*}

The results are illustrated in Fig. \ref{fig:wd_back}, where in the left panel we show WD-fTDEs background from dwarf galaxies, while in the right panel the WD-fTDEs background from GCs. As done in the case of the MS background, we compute the SNR for these scenarios as well. In all cases, the SNR remains below 1, with the exception of the GWB produced by WD-fTDEs in dwarf galaxies. For this specific signal, the SNR for DO is approximately 19, assuming a 10-year observation period.  \footnote{\textcolor{black}{Note that, while calculating the detection rates and background contributions of WD-fTDEs, we assume an occupation fraction of 1 for both globular clusters and dwarf galaxies. This assumption is further discussed in Appendix \ref{app:of}. In Appendix \ref{app:lighter}, we also explore the case where the BH hosted within the GC is lighter.}}

\section{Repeated partial disruptions}
\label{sec:rptdes}
In this Section we cover rpTDEs, which occur under the empty loss cone regime, i.e. when $q<1$ (cf. Section \ref{sec:ftdes}). 

We anticipate that our results show that even in an (unrealistic) optimistic scenario - with all the bursts from each TDE above threshold - the rate of detections is below one in 10 years. Nevertheless, we outline the methodology used to derive these rates, since it can be a useful technique also for other transient sources.

\subsection{Individual detections}
We consider four rpTDE populations, created with the methodology described in \citealt{Broggi:24aa}. In each population we have Sun-like stars, disrupted by four different BH sizes: $M_1=10^4\text{M}_{\odot}$, $M_2=3\times 10^5\text{M}_{\odot}$, $M_3=4\times 10^6\text{M}_{\odot}$ and $M_4= 10^7\text{M}_{\odot}$. 
In each scenario, we assume that the $j$-th rpTDE is characterized by a constant period $P_j$ between subsequent passages. This means that each rpTDE is modeled as a series of periodic bursts. 
\footnote{In reality, $P_j$ evolves with the orbital energy, but assuming it constant serves as a reasonable first approximation.}

Let $\Delta t_{\rm obs}$ represent the detector observation time, assumed to be 10 years. There are two possible scenarios:
\begin{enumerate}
    \item long period, $P_j \geq \Delta t_{\rm obs}$ : the detector observes only a single burst (the first or any subsequent one) from the $j$-th rpTDE, and no SNR accumulation occurs;  
    \item short period, $P_j <\Delta t_{\rm obs}$: more consecutive bursts produced by the $j$-th rpTDE accumulate at the detector.  
\end{enumerate}

To address this, we begin by considering an rpTDE characterized by an initial specific orbital energy $\epsilon_j$. The disruption rate for events within the energy range $[\epsilon_j, \epsilon_j +  d\epsilon_j]$ is given by 
\begin{align}
\Gamma_j = d\epsilon_j \, \frac{\partial \dot{N}}{d\epsilon_j} \,.
\end{align}
The $i$-th passage from the $j$-th rpTDE occurred at 
\begin{align}
t_{\rm now} - t_{\rm travel} - (i-1) P_j \, ,
\end{align}
where $t_{\rm travel}$ is the time the signal needs to travel from the event to the detector. This leads us to assign a weight $w$ to the event (detection of the $i$-th burst or a sequence of bursts starting with the $i$-th), expressing the relative probability of observing such an event within $\Delta t_{\rm obs}$.
 
In general, since the event has been produced between $t_1 = t_\mathrm{now} - t_\mathrm{travel}-t^i_0$ and $t_2 = t_\mathrm{now} - t_\mathrm{travel}-t^i_f$, we define the corresponding relative weight as 
\begin{equation}
    w\equiv \frac{\int_{t_1}^{t_2} \, \Gamma_j \, dt'}{\int_{0}^{\Delta t_\mathrm{obs}} \, \Gamma_j \, dt'} = \frac{t^i_f - t^i_0}{\Delta t_\mathrm{obs}}
\end{equation}
and it can be interpreted as the fraction of relevant events.

\subsection{Long period}
The rate of detected rpTDEs will be
\begin{align}
    \dot{N}^{\rm rpTDEs}_{\rm det}=&\int dz \int d\epsilon \,\frac{d\dot{N}^{\rm rpTDEs}_{\rm det}}{dzd\epsilon}=\\\nonumber
    &\int dz \frac{4\pi c \chi^2}{H(z)}\times\frac{\Phi(z)}{1+z}\int d\epsilon\frac{d\dot N}{d\epsilon}\times\Theta(\epsilon,z).
\end{align}
$\Phi(z)$ is the distribution of BHs through the Universe, assuming a fixed BH mass among the four listed above. It is determined by integrating the BH mass over a specific mass bin, considering this value as a representative estimate,
\begin{align}
  \Phi(z)= \Pi(a,b,M_\bullet) \int_{a}^{b} dM_\bullet \Phi({M}_\bullet , z),
\end{align}
where 
\begin{align}
\Pi(a,b,M_\bullet)=\Theta(M_\bullet -a)\Theta(b-M_\bullet)=\begin{cases}
    1 \,\,\,\, a\leq M_\bullet \leq b,\\
    0 \,\,\,\, \text{else}.
\end{cases}
\end{align}
Each mass bin is taken centered on the BH mass considered.

The integral over the specific orbital energy can be written as a discrete sum over the energy bins
\begin{align}
 \sum_{j}\Delta \epsilon_{j} \frac{d\dot{N}}{d\epsilon_j}\Theta_j= \sum_{j}\Delta \epsilon_{j} \frac{d\dot{N}}{d\epsilon_j}\sum_{i}^{N_{\rm bursts}(\Delta\epsilon_j)}\Theta_i.   
\end{align}
Since the detector can potentially see the first or any subsequent bursts from the $j$-th rpTDE, the corresponding $\Theta_j$ will range from 0 to $N_{\rm bursts}(\Delta\epsilon_j)$, depending on how many bursts are above the threshold.

\subsection{Short period}
When the observational time is longer than the time between subsequent passages, the detector will collect signal by more passages of the same star, so that the strain of the event will accumulate.
We first define the quantities $\tilde{n}_j$ and $p_j<P_j$ as
\begin{equation}
    \Delta t_\mathrm{obs} = \tilde{n}_j\, P_j + p_j
\end{equation}
where $\tilde{n}_j$ represents the maximum (integer) number of subsequent signals that the detector can accumulate. In particular the detector can receive
\begin{enumerate}
    \item  the first $1 \leq i\leq \tilde{n}_j$ signals, each with weight $w_{\rm \alpha} = P_j/\Delta t_\mathrm{obs}$;
    \item a set of $\tilde{n}_j+1$ subsequent signals, each with weight $w_{\rm \beta} = p_j/\Delta t_\mathrm{obs}$;
    \item a set of $\tilde{n}$ subsequent signals, each with weight $w_{\rm \gamma} = (P_j-p_j)/\Delta t_\mathrm{obs}$;
    \item the last $i$ signals, with $1\leq i \leq \tilde{n}_j$, each with weight $w_{\rm \delta} = P_j/\Delta t_\mathrm{obs}$.
\end{enumerate}

To better understand these categories, let us consider a practical example, where we have $N_{\rm bursts}=5$
and $P_j= 3$ years. Then we have 
\begin{align}
    \tilde{n}=3,\,\,\, p= 1\,\text{year}.
\end{align}
Calling $n_j=\Gamma_j \Delta t_{\rm obs}$ the number of rpTDEs generated at energy $E_{j}$, the detector can observe the following combinations of bursts with corresponding weights:
\begin{itemize}
    \item passage 1; total number of events: 0.3$n_j$; 
    \item passage 1 and 2; total number of events: 0.3$n_j$;
    \item passage 1 and 2 and 3; total number of events: 0.3$n_j$;
    \item passage 1 and 2 and 3 and 4; total number of events: 0.1$n_j$;
    \item passage 2 and 3 and 4; total number of events: 0.2$n_j$;
    \item passage 2 and 3 and 4 and 5; total number of events: 0.1$n_j$;
    \item passage 3 and 4 and 5; total number of events: 0.3$n_j$;
    \item passage 4 and 5; total number of events: 0.3$n_j$;
    \item passage 5; total number of events: 0.3$n_j$.
\end{itemize}

For the detection rate, the main difference from the case described in the previous paragraph lies in the way $\Theta_j$ is expressed. Here $\Theta_j$ is the linear combination of the weighted $\Theta_i$ corresponding to the different combinations of passages. Considering the example before, $\Theta_j$ will range among 0 and
\begin{align}
    \Theta_j = &w_{\alpha}\Theta_1+w_{\alpha}\Theta_2+w_{\alpha}\Theta_3+w_{\beta}\Theta_4+w_{\beta}\Theta_5+w_{\gamma}\Theta_6 \nonumber \\
   &+w_{\delta}\Theta_7+w_{\delta}\Theta_8+ w_{\delta}\Theta_9 =1+ (N_{\rm burst} -1)\frac{P_j}{\Delta t_{\rm obs}} \approx 2.2,
\label{eq:theta_j}
\end{align}
if all the combination of bursts are above the SNR threshold.
This occurs because if the root sum square SNR of a specific combination of passages is above a certain threshold, then the corresponding $\Theta_i$ is set to 1, but it is still multiplied by a weight (less than 1). Thus,
Eq.~\eqref{eq:theta_j} states that the total $\Theta$ for the j-th rpTDE is the weighted sum of the $\Theta_i$ values for each valid burst combination of that rpTDE.
The theoretical upper limit corresponds to the case in which all $\Theta_i$ are equal to 1, so the total becomes the sum of the weights for that rpTDE.

Our calculation shows that none of the detector configurations under consideration will observe individual detections of these events.
\begin{figure}
        \centering
        \includegraphics[width=0.5\textwidth]{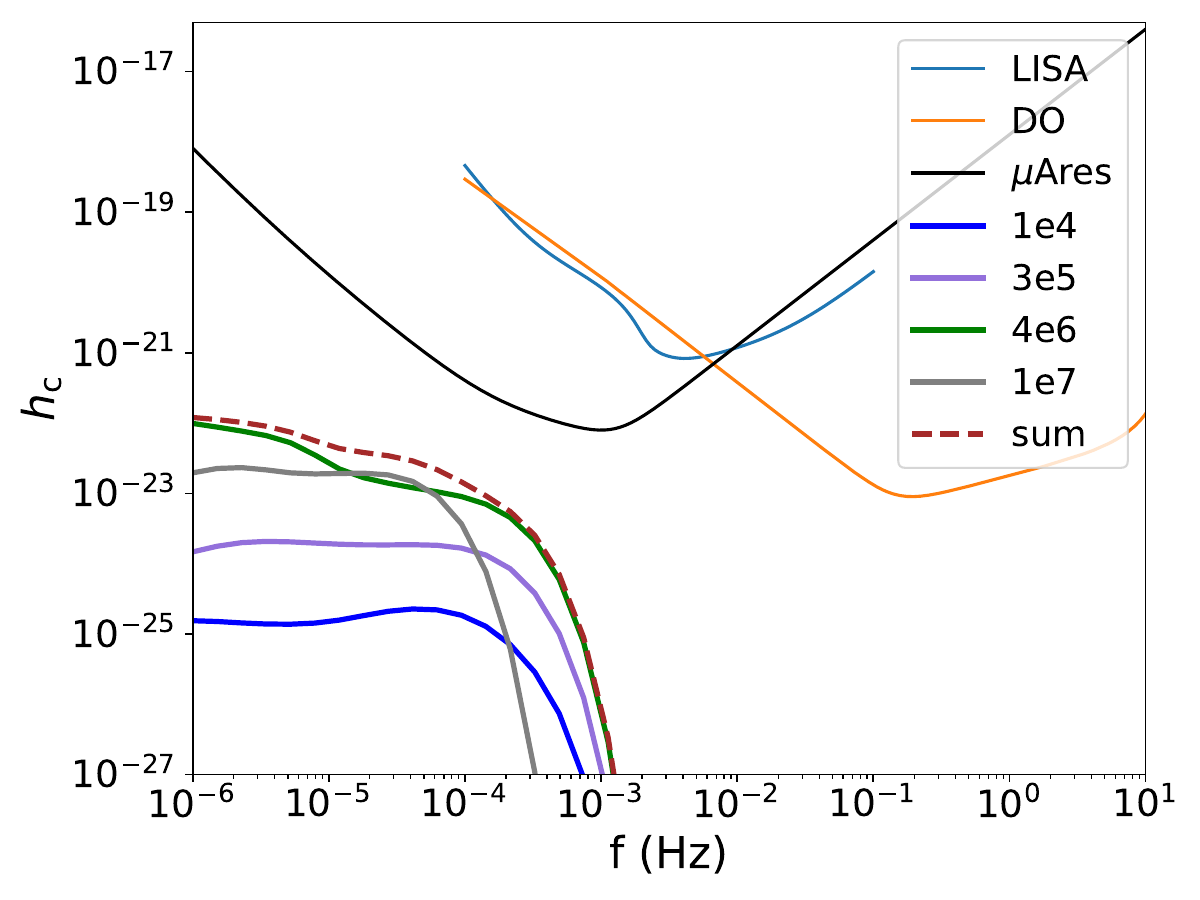}
        \caption{Background signals from different populations of rpTDEs: the solid-blue line is the signal from rpTDEs with a BH of mass $M_\bullet= 10^4\text{M}_{\odot}$, the solid-purple line with $M_\bullet=3\times 10^5\text{M}_{\odot}$, the solid-green line with $M_\bullet=4\times 10^6\text{M}_{\odot}$ and the solid-gray line with $M_\bullet=10^7\text{M}_{\odot}$. The dashed brown line is the sum of all the backgrounds.}
        \label{fig:back_rptde}
\end{figure}

\subsection{The background}

The background from rpTDEs is 
\begin{align}
    h^2_{\rm {c, pop}}=\frac{G}{c^3\pi^2}\times \frac{1}{f}&\sum_{l}\Delta z_{l} \frac{4\pi c\chi^2(z_l)}{H(z_l)}\times\frac{\Phi(z_{l})}{(1+z_{\rm l})\chi^2(z_l)}\\\nonumber
    &\times\sum_{j}\Delta \epsilon_{j} \frac{\Delta\dot{N}}{\Delta\epsilon_j}\sum_{i}^{N(\Delta\epsilon_j)}\frac{dE_i}{df}.
\end{align}
with $dE_i/df$ being the energy spectrum (Eq.~\ref{eq:energy_spectrum}) from the  $i$-th burst. So, contrary to the previous case for individual detections, here we sum over all passages without considering specific combinations of bursts.

The results are shown in Fig.~\ref{fig:back_rptde}, where the signals corresponding to different BH masses are represented by solid lines in various colors ($M_1=10^4\text{M}_{\odot}$ - blue, $M_2=3 \times 10^5\text{M}_{\odot}$ - red, $M_3=4\times 10^6\text{M}_{\odot}$ - purple, and $M_4= 10^7\text{M}_{\odot}$ - gray). The combined signal, which includes contributions from all these masses, is depicted with a dashed brown line. Notably, the most significant contribution, dominating over the others, comes from the population with $M_{\bullet}=4\times 10^6\text{M}_{\odot}$, that is the case where there is the highest number of TDEs around $q=1$. For higher BH masses, since all TDEs are distributed around values of $q$ strictly smaller than one, the total GW emission is becoming less relevant. In fact, systems with a larger $M_\bullet$ produce TDEs exclusively in the empty loss cone regime. While this increases the number of subsequent partial disruptions, the small values of $q$ ($\lesssim 10^{-3}$) imply that the pericenter at which the sequence of partial disruptions occurs is so large, corresponding to lower frequency GW bursts. This pushes the signal outside of the LISA band, as clearly demonstrated by the corresponding contribution to the GWB shown in Fig.~\ref{fig:back_rptde} by the gray line. The SNR for this background from rpTDEs is below 1 for all the three instruments considered.
\begin{figure*}
    \centering
    \begin{minipage}{0.48\textwidth}
        \centering
        \includegraphics[width=\linewidth]{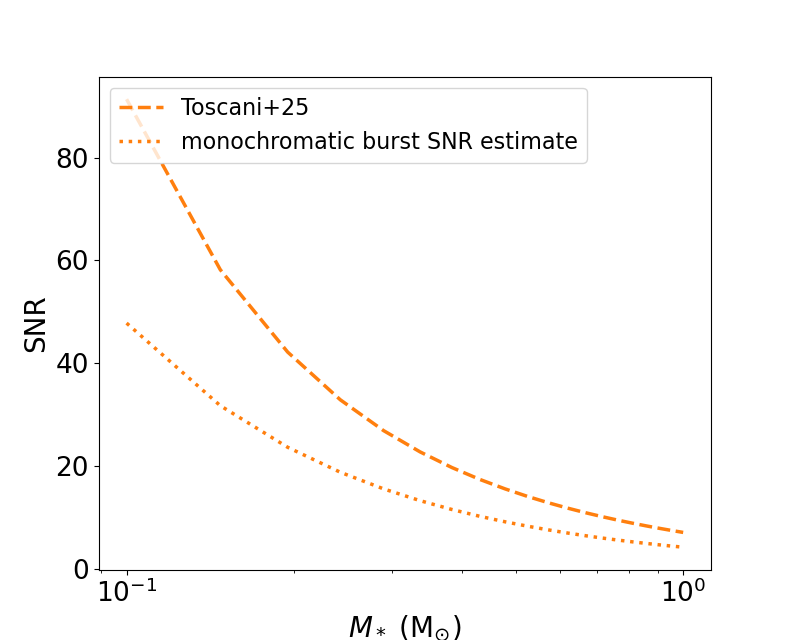}
        \label{fig:snr_mstar}
    \end{minipage}
    \hfill
    \begin{minipage}{0.48\textwidth}
        \centering
        \includegraphics[width=\linewidth]{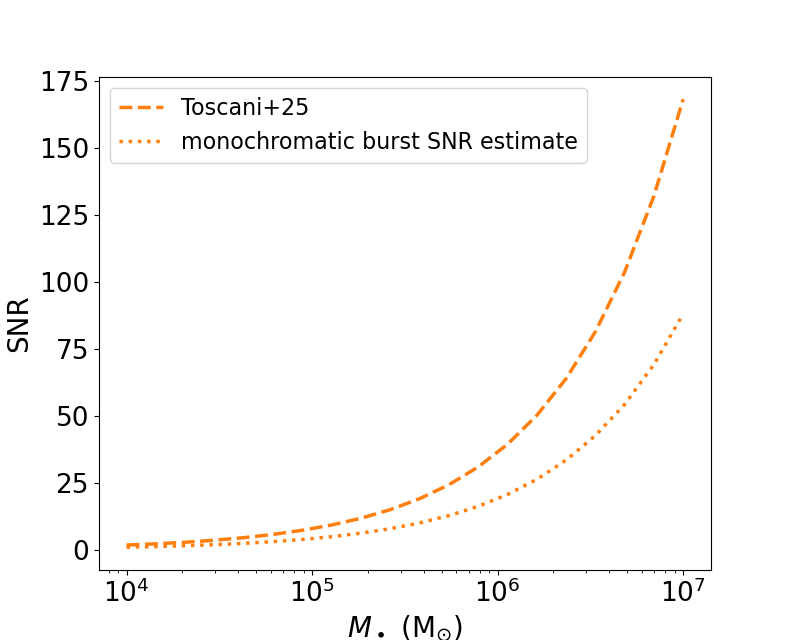}
        \label{fig:snr_bh}
    \end{minipage}
    \caption{SNR computed for the DO instrument considered in this work. Left: SNR as a function of stellar mass for a fTDE around a $4 \times 10^6\text{M}_{\odot}$ BH at 8 kpc. Right: SNR as a function of BH mass for a fTDE of a $0.1\text{M}_{\odot}$ MS star, also at 8 kpc. In both panels, the dashed line represents the SNR from Eq.~\eqref{eq:snr}, while the dots show the SNR estimates based on the monochromatic burst approximation (cf.~\citealt{Pfister:22aa}).}
    \label{fig:snr_combined}
\end{figure*}

\begin{figure}
        \centering
        \includegraphics[width=\linewidth]{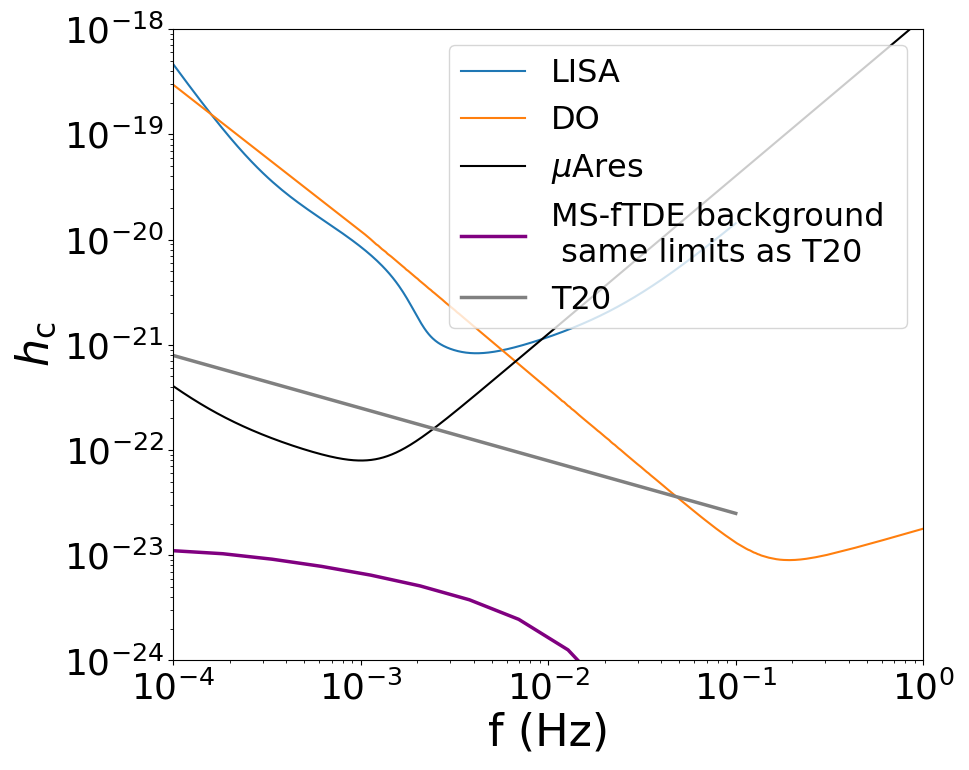}
        \caption{GW background from a population of MS-fTDEs, calculated with the same integration limits as in \citealt{Toscani:20aa} (violet curve). The gray line represents the original signal as in \citealt{Toscani:20aa}.}
        \label{fig:reproduction_toscani2020}
\end{figure}
\begin{figure*}
    \centering
    \begin{minipage}{0.33\textwidth}
        \centering
        \includegraphics[width=\linewidth]{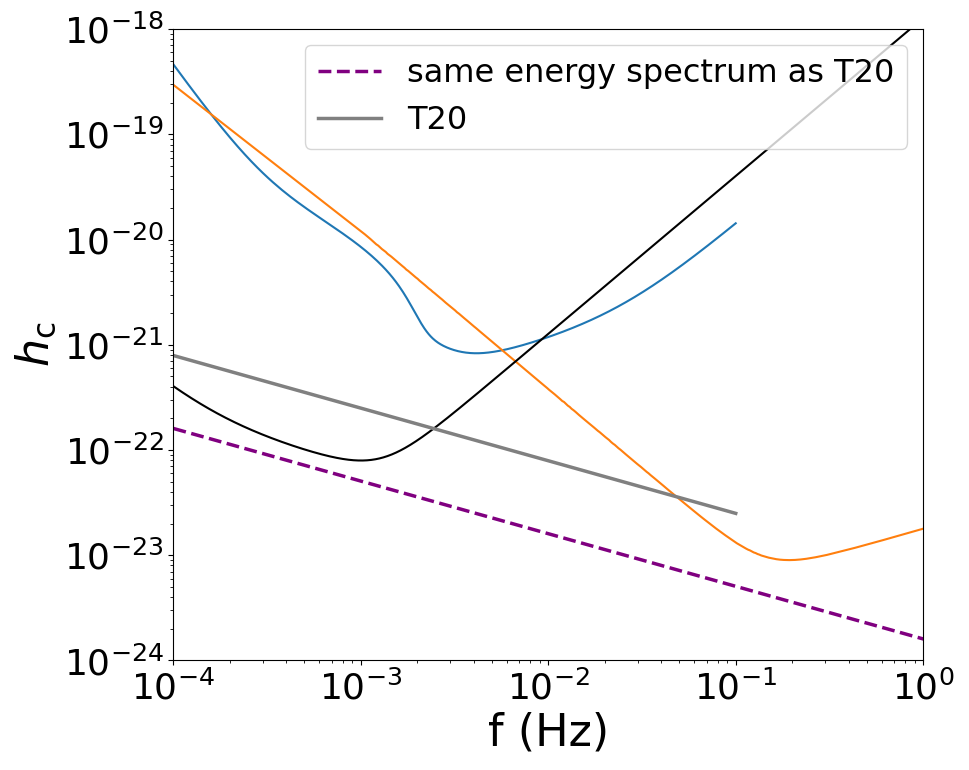}
    \end{minipage}
    \hfill
    \begin{minipage}{0.33\textwidth}
        \centering
        \includegraphics[width=\linewidth]{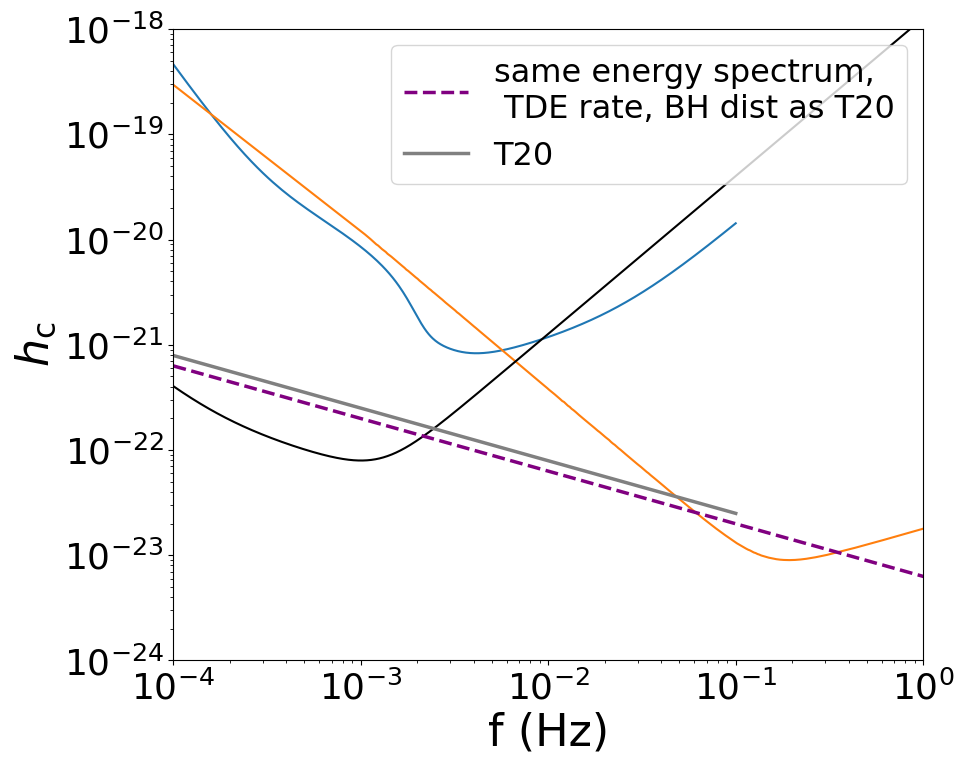}
    \end{minipage}
    \hfill
    \begin{minipage}{0.33\textwidth}
        \centering
        \includegraphics[width=\linewidth]{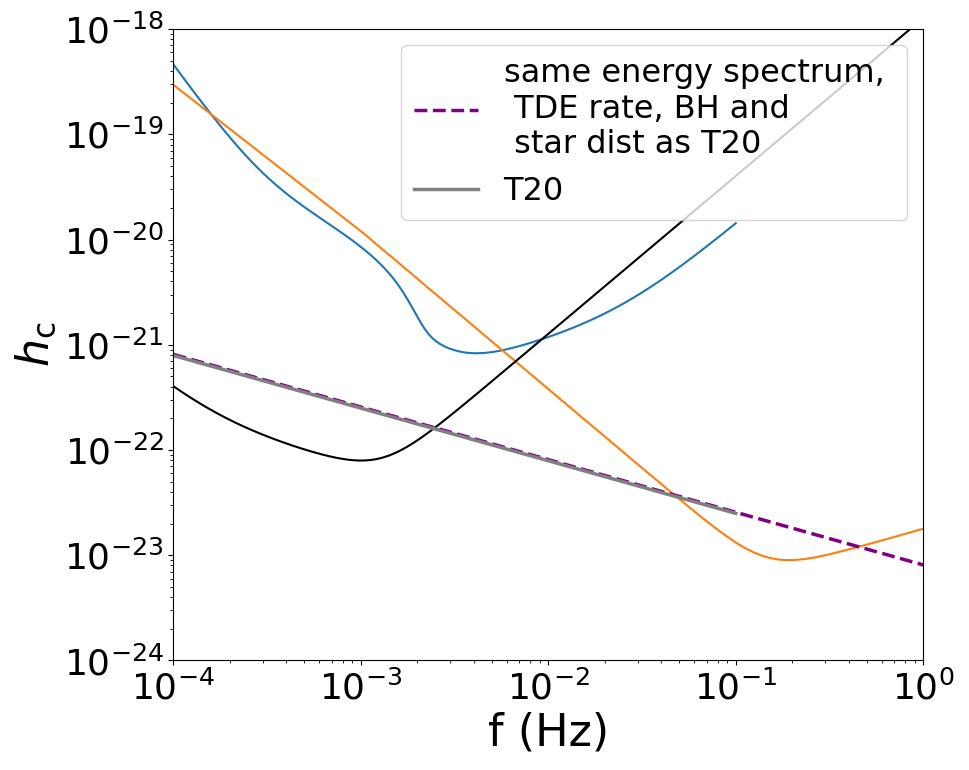}
    \end{minipage}
    \caption{Steps to reproduce the signal from T20. Left panel: use of the same energy spectrum. Central panel: in addition to the previous change, use of the same rate and BH mass function. Right panel: in addition \textcolor{black}{to} the previous changes, use of the same stellar mass function.}
    \label{fig:reproduction_toscani2020_2}
\end{figure*}

\begin{figure*}
    \centering
    \begin{minipage}{0.33\textwidth}
        \centering
        \includegraphics[width=\linewidth]{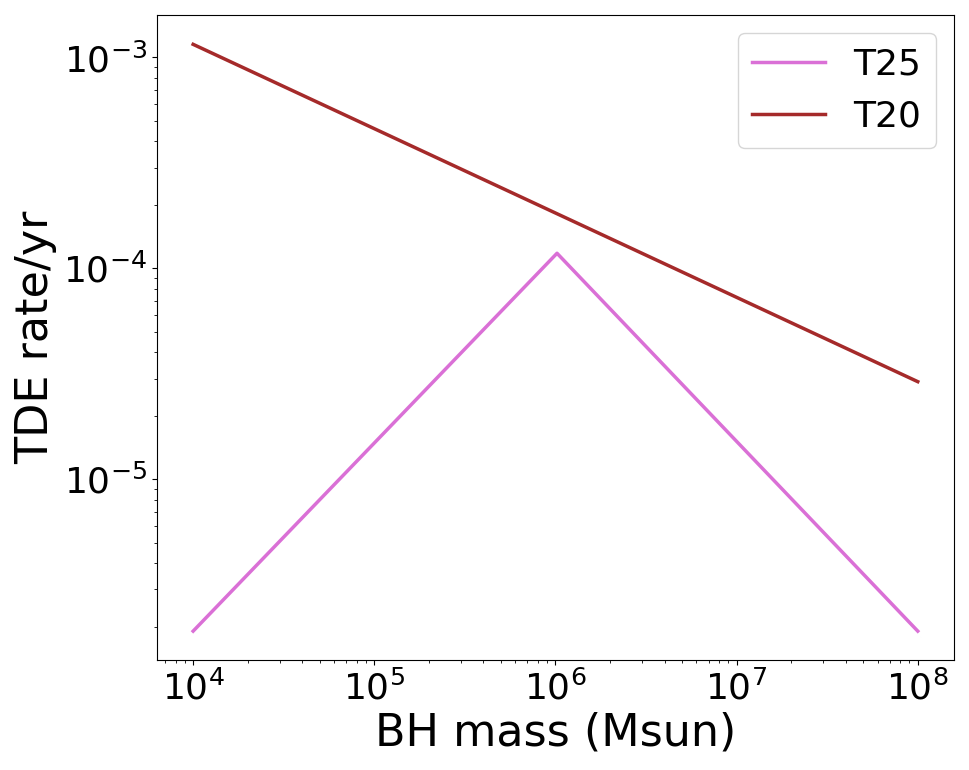}
    \end{minipage}
    \hfill
    \begin{minipage}{0.33\textwidth}
        \centering
        \includegraphics[width=\linewidth]{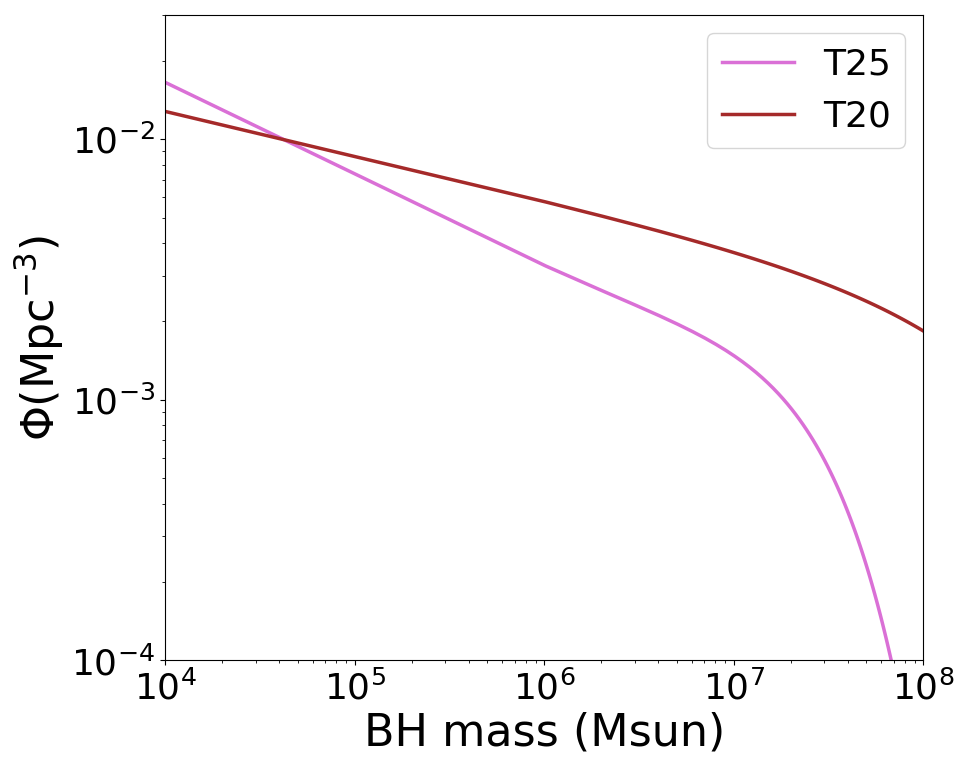}
    \end{minipage}
    \hfill
    \begin{minipage}{0.33\textwidth}
        \centering
        \includegraphics[width=\linewidth]{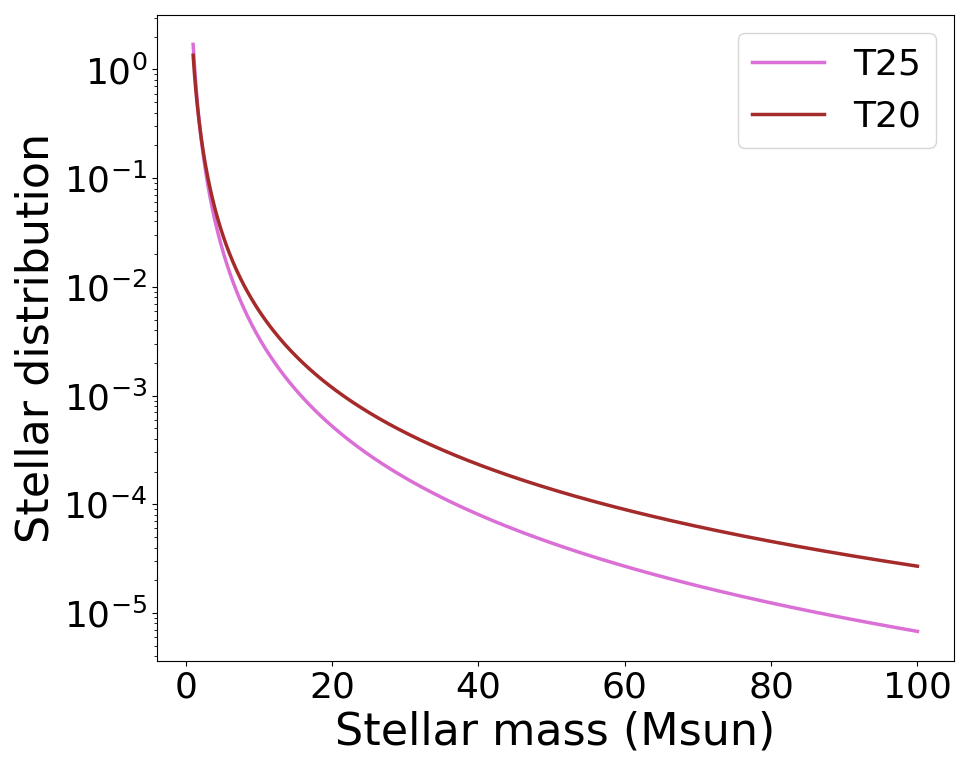}
    \end{minipage}
    \caption{Left panel: TDE rate as a function of BH mass. Central panel: BH mass function (assuming a fixed redshift of $z=0.1$). Right panel: Stellar mass function. In yellow we show the functions used in T20, while in purple the functions used in this current work.
    }
    \label{fig:reproduction_toscani2020_3}
\end{figure*}

\section{Discussion}
\label{sec:discussion}
Our study shows that the detection of rpTDEs with GWs will be unlikely in the near future. This is related to the typical values of the penetration factor that characterize these events. The penetration factor is a dimensionless quantity defined as 
\begin{align}
    \beta = \frac{R_{\rm t}}{R_{\rm p}},
\end{align}
which, in the empty loss cone regime,  is smaller than 1. In fact, two-body relaxation is the main mechanism for a star on a given orbit to change its orbital parameters. Quantitatively, a star with pericenter comparable to its tidal radius will stay on average $1/q$ periods on the same orbit before changing significantly its angular momentum \textcolor{black}{\citep{Bortolas:23aa}}. When $q$ is small, the relaxation is very slow and stars will gradually increase their penetration factor from $\sim0$ up to a critical value such that in $\sim1/q$ orbits the star is completely destroyed because of the subsequent partial disruptions happened. Since the mass lost at each passage is smaller when $\beta$ is smaller, a general trend emerges: the smaller $q$, the smaller the mass lost at each passage, the smaller the corresponding penetration factor. A direct consequence is that, in a given galaxy, the greatest GW emission of repeating partial disruptions corresponds to energies such that $q\to1^-$, i.e. systems with a number of repetitions of the order of unity as these stars will emit their bursts close to the tidal radius. Systems with mass $M_\bullet \gtrsim4\times10^6 \text{M}_\odot$ produce the vast majority of their TDEs where $q\lesssim 10^{-2}$, meaning that the pericenter of the corresponding rPTDEs is large. Because of the larger pericenter, empty loss cone sources
result in GW emission peaked around lower frequencies than the GWs emitted by TDEs in the full loss cone scenario. In such a low frequency window, future instruments are not sensitive enough to observe them, and also the accumulation of SNR due to multiple passages of the star around the BH is not enough to overcome this problem. 

As for the fTDE scenario instead, the situation seems more promising, as we analyze in the following.

\subsection{Individual detections of fTDEs}
\subsubsection{MS-fTDEs: SNR}
We start our analysis by examining the impact of including harmonics on the SNR of TDEs. In the left panel of Fig.~\ref{fig:snr_combined} we plot the SNR for a MS-fTDE with a fixed BH mass of $4\times 10^6\text{M}_{\odot}$ at a distance of 8 kpc. We consider stellar masses in the range $0.1 \text{M}_{\odot}\leq M_* \leq 1\text{M}_{\odot}$, each disrupted at a pericenter equal to their tidal raidus and, for illustrative purposes, we only consider the DO detector. The dashed line represents the SNR computed using Eq.~\eqref{eq:snr}, while the dots show the SNR estimates following the prescription of \citealt{Pfister:22aa} (hereafter P22), which models the GW signal from a TDE as a monochromatic burst. The curve that includes the contribution from harmonics is consistently higher, with the discrepancy becoming more relevant at lower stellar masses, reaching up to a factor of $\sim$2. Additionally, we observe an interesting trend: the SNR decreases with increasing stellar mass. Although this may seem counterintuitive, it results from the signal peak shifting toward a less favorable frequency range as the stellar mass increases. 

In the right panel of the same figure, we consider instead the SNR of a  0.1$\text{M}_{\odot}$ MS-star at 8 kpc, destroyed  by a BH of mass ranging in the interval $10^4\text{M}_{\odot}\leq M_\bullet \leq 10^7\text{M}_{\odot}$. As in the previous case, the estimate that includes all harmonic contributions yields a higher SNR, with differences reaching up to a factor of $\sim$2. However, in contrast to the earlier trend, the SNR ratio increases with black hole mass—this is because the signal reaches a higher peak for more massive black holes. 

\subsubsection{MS-fTDEs: detection rates}
If we compare our results (Fig. \ref{fig:MS_ftde}) with those presented in P22 (Fig. 10)\footnote{The results in Fig. 10 of P22 are detections over 1 year. So, in order to have detection over 10 years, they need to be multiplied by a factor 10.},  our detection rates are lower. This discrepancy is first due to the lower TDE rate adopted in this work (cf. Fig. 8 from P22 and left panel of Fig.\ref{fig:reproduction_toscani2020_3} below). However, since our chosen TDE rate is based on observational constraints, we consider it a reliable estimate. Then, the detector curves that we use in this work are also different. The LISA sensitivity adopted in this study is more accurate since we also include the noise from galactic WD binaries, 
which in contrast makes our curve less optimistic than the one used in P22, especially in the frequency window where the signal from MS-fTDEs peaks.  For DO, the curve we use falls between the ALIA and DECIGO/BBO sensitivities presented in P22. Thus, it is not possible to do an exact one-to-one comparison. Nevertheless, our findings are consistent with the main prediction of that study, i.e. the detectors operating in the dHz range are more likely to observe these events.

\subsubsection{WD-fTDEs}
As for WD-fTDEs, from Fig.~\ref{fig:wd_individual_det} we conclude that we have no possible detection foreseen for LISA. However, if we consider the DO detector, the situation becomes more optimistic. Indeed, it could  marginally see up to a few tens of WD-fTDEs from GCs. This is still a very uncertain number, mainly due to the strong constraints we put on our estimates (fixed BH mass of one thousand solar masses in all the GCs and occupation fraction of 1), but given the uncertainty surrounding the presence of IMBHs in these environments, we could not adjust these estimates any further. The detection rates instead increase to hundreds over a 10 year-time observation if we consider WD-fTDEs from dwarf galaxies, with SNR up to $\sim 45$. With respect to MS stars, this scenario is clearly favored due to the typical peak frequencies of these signals, $\sim 0.1$ Hz, which fall around the most sensitive region of the detector.

\subsection{The background of fTDEs}
Let us now analyze the MS-background signal. As anticipated in Sec.~\ref{sec:ftdes_ms}, a direct one-to-one comparison between our results and those of \citealt{Toscani:20aa} (T20 from now on) is challenging due to differences in the integration limits. Therefore, our first step is to reproduce the results of T20 considering the same regime of integration, while employing our current prescription for the background. The outcome of this comparison is illustrated in Fig.~\ref{fig:reproduction_toscani2020}, where we observe that, in the frequency range where the T20 signal is represented, our signal differs by approximately two orders of magnitude (before the steep fall at $0.01$ Hz).

To understand the origin of this discrepancy, we implement the following modifications step by step. First of all, the limits used in T20 are the following: $0\leq z \leq 3$, $1\text{M}_{\odot}\leq M_* \leq 100\text{M}_{\odot}$, $10^{6}\text{M}_{\odot}\leq M_\bullet \leq 10^8\text{M}_{\odot}$ and $1\leq \beta \leq \beta_{\rm max}(M_\bullet, M_*)$. Note that in the work of T20 they consider the dimensionless parameter $\beta$ instead of the pericenter, with $\beta_{\rm max}$ calculated assuming the pericenter equal to the Schwarzschild radius of the BH. 

First, instead of using the energy spectrum described in Eq.~\eqref{eq:energy_spectrum}, we adopt the version presented in T20. This alternative spectrum differs mainly by a constant factor of $\sim 80$, which arises from a different approximation used in T20, as well as from the omission of harmonics. The exclusion of harmonics restores the characteristic frequency dependence $f^{-0.5}$, which is a key result highlighted in T20. Furthermore, the change in the constant factor leads to an enhancement of the signal by roughly one order of magnitude (Fig.~\ref{fig:reproduction_toscani2020_2}, left panel). 

Next, employing the same TDE rate and BH mass function as in the previous work, we observe an additional increase by a factor of $\sim 5$ (Fig.~\ref{fig:reproduction_toscani2020_2}, central panel). Finally, we modify the stellar mass function. Instead of using the function defined in Eq.~\eqref{eq:stellar_mass_function}, normalized within the mass range (1\text{M}$_{\odot}$, 100\text{M}$_{\odot}$), we use the normalised stellar mass function as in T20, $ \propto M_*^{-2.3}$ (Fig.~\ref{fig:reproduction_toscani2020_2}, right panel). Thus, we recover the results from T20. 

While the difference arising from our chosen energy spectrum is inherent to the motivation of our work and cannot be altered, we illustrate in Fig.~\ref{fig:reproduction_toscani2020_3} the TDE rate, BH mass function, and stellar mass function used in T20 and in this study. The most significant discrepancy lies in the TDE rate, which differs by more than two orders of magnitude at both low and high BH masses. The BH mass function also shows a notable difference, particularly for BH masses above 10$^6$M$_{\odot}$. Similarly, the stellar mass function adopted in T20 assigns greater weight to more massive stars. However, our choice of TDE rate is guided by observation constraints, and the stellar mass function used in this work is more physically reasonable, therefore, we regard the current, more conservative results as reliable and an updated version of the results of T20. Yet, this results in a GWB with SNR smaller than 1 even if we assume 10 years of observation for each of the three instruments.

As for the WD-fTDE background, we have that the signal produced within GCs is found to be undetectable. In contrast, the GWB generated by WDs disrupted by intermediate-mass BHs in dwarf galaxies appears much more promising: with an SNR of approximately 19 over a 10-year observation period, it falls well within the sensitivity range of DO. Thus, these TDEs offer one of the most promising avenues for probing the existence of the elusive population of intermediate-mass BHs, for which observational evidence remains extremely limited to date.

\section{Conclusions}
\label{sec:conclusions}
In this work, we have computed the detection rates of TDEs for two future space-based GW detectors: LISA and DO. Compared to previous studies, our analysis benefits from more realistic sensitivity curves and improved modeling, including the full contribution of all harmonics in the energy spectrum. 

For stars fully disrupted in a single passage, our results, although more conservative, are in agreement with previous studies and they confirm a key finding: these sources are promising targets for dHz observatories, but unlikely to be detectable by LISA.

Among the most compelling targets are WDs disrupted by intermediate-mass BHs in dwarf galaxies. These events stand out both as individually detectable sources — with detection rates over 10 years ranging from tens to hundreds depending on the SNR threshold considered — and as the only class producing a background potentially strong enough to be observed (SNR $\sim 19$ over a 10-year period of observation).

Furthermore, for the first time, we have considered the GW emission from repeated disruptions, considering two scenarios: the case where the period between each passage is longer than the detector lifetime and the case where it is shorter, potentially leading to an accumulation of bursts (and SNR) at the detector. We find that the largest signal comes from systems with a BH mass around $4\times10^6\ \text{M}_\odot$. Systems with less massive BHs produce only a small number of rpTDEs, and overall their GW signal decreases. Systems with larger BHs, on the other hand, predominantly produce rpTDEs, but at pericenters too large to emit detectable GW signals - even considering the accumulation of multiple passages. Unfortunately, our findings show that the detection of these events with GWs is unlikely in the near future.

With this work, we provide updated estimates of the GW emission from TDEs, and we hope that this can be a useful reference to drive the development of deci-Hz detectors. 

\begin{acknowledgements}
MT, LB and AS acknowledge financial support provided under the European Union’s H2020 ERC Consolidator Grant “Binary Massive Black Hole Astrophysics” (B Massive, Grant Agreement: 818691). MT acknowledges N. Tamanini for useful feedback. MT acknowledges N. Cuello for continuous support.  
\end{acknowledgements}

\bibliographystyle{aa}
\bibliography{aa55648-25}

\begin{appendix}
\onecolumn
\section{Occupation fraction}
\label{app:of}
\begin{figure*}[h]
    \centering
    \begin{minipage}{0.49\textwidth}
        \centering
        \includegraphics[width=\linewidth]{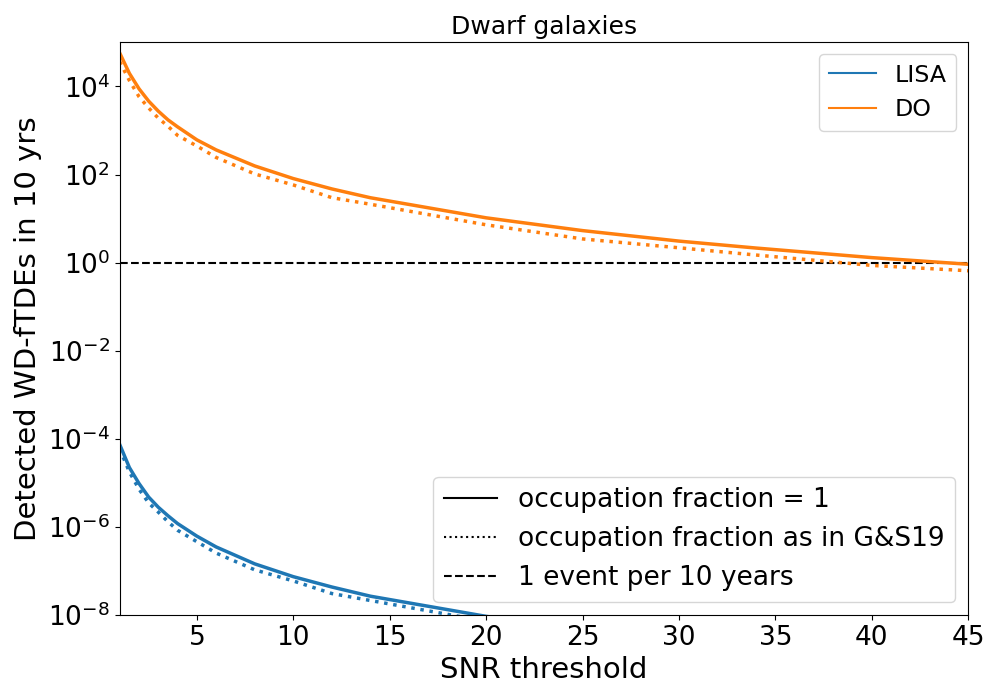}
    \end{minipage}
    \hfill
    \begin{minipage}{0.49\textwidth}
        \centering
        \includegraphics[width=\linewidth]{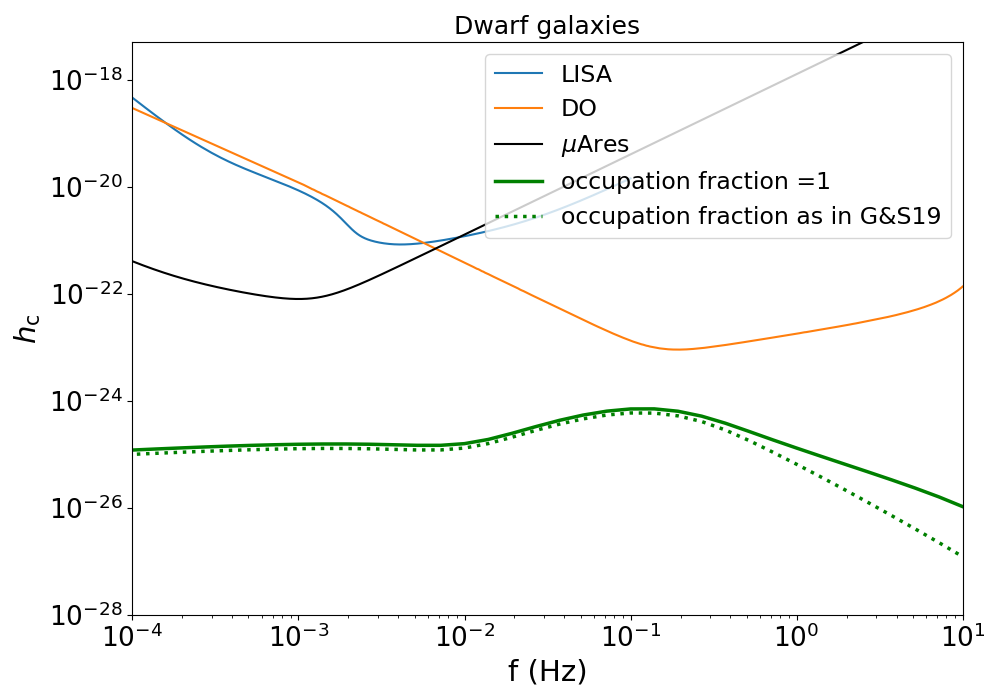}
    \end{minipage}
    \caption{\textcolor{black}{Left panel: detection rates of WD-fTDEs from dwarf galaxies plotted for different SNR thresholds. Right panel: GWB from WD-fTDEs from dwarf galaxies. The color scheme and the detectors are the same as in Fig. \ref{fig:wd_individual_det}-\ref{fig:wd_back}. The solid line represents an occupation fraction of 1, while the dotted line an occupation fraction as in \citealt{Gallo:19aa}.}}
    \label{fig:occfrac_WDFTDEs}
\end{figure*}

\textcolor{black}{While calculating the WD-fTDEs detection rates and background for dwarf galaxies, we considered the same BH mass function as in \citealt{Pfister:22aa}, which assumes an occupation fraction of 1. While this is in general true for supermassive BHs in galactic cores, the situation for IMBHs is likely different and dependent on the galaxy mass. To account for this, we repeated the calculation using an occupation fraction that depends on the stellar mass of the host galaxy. This factor enters Eq. \eqref{eq:tde_diff_rate} as a multiplicative term modifying the BH mass function. Specifically, we implemented the parametrization proposed by \citealt{Gallo:19aa},
\begin{align}
    \lambda_{\rm occ}(M_{\rm gal})= 0.5 +0.5*\tanh{\left(2.5^{|8.9-\log{M_{\rm gal, 0}/M_{\odot}}|}\log\frac{M_{\rm gal}}{M_{\rm gal, 0}}\right)},
\end{align}
with $M_{\rm gal}$ galaxy stellar mass and $M_{\rm gal, 0}$ parameter that we assume equal to $10^{8.17}\text{M}_{\odot}$ (cf. Fig. 1 of \citealt{Gallo:19aa}). $M_{\rm gal}$ can be converted in the BH mass using the relation from \citealt{Reines:15aa}.\newline\indent
The results obtained using this revised occupation fraction are shown in Fig. \ref{fig:occfrac_WDFTDEs}. The left panel presents the individual detection rates, while the right panel displays the background. Although some variations are noticeable, they do not significantly impact the conclusions discussed in the main body of the paper. Given this, and considering the current uncertainties surrounding the distribution of IMBHs in dwarf galaxies, we regard the results in the main text as a valid upper limit.\newline\indent For GCs, the situation is even more uncertain. To explore the impact on detection rates, we simply assume a constant occupation fraction $\lambda_{\rm occ}$. This parameter acts as a multiplicative factor on the TDE rate per cluster, while the GWB scales with $\sqrt{\lambda_{\rm occ}}$. Also in this scenario, given the significant uncertainty associated with this parameter, we continue to treat the results presented in the main text as a valid upper limit.}

\section{Lighter IMBH}
\label{app:lighter}
\begin{figure*}[h]
    \centering
    \begin{minipage}{0.49\textwidth}
        \centering
        \includegraphics[width=\linewidth]{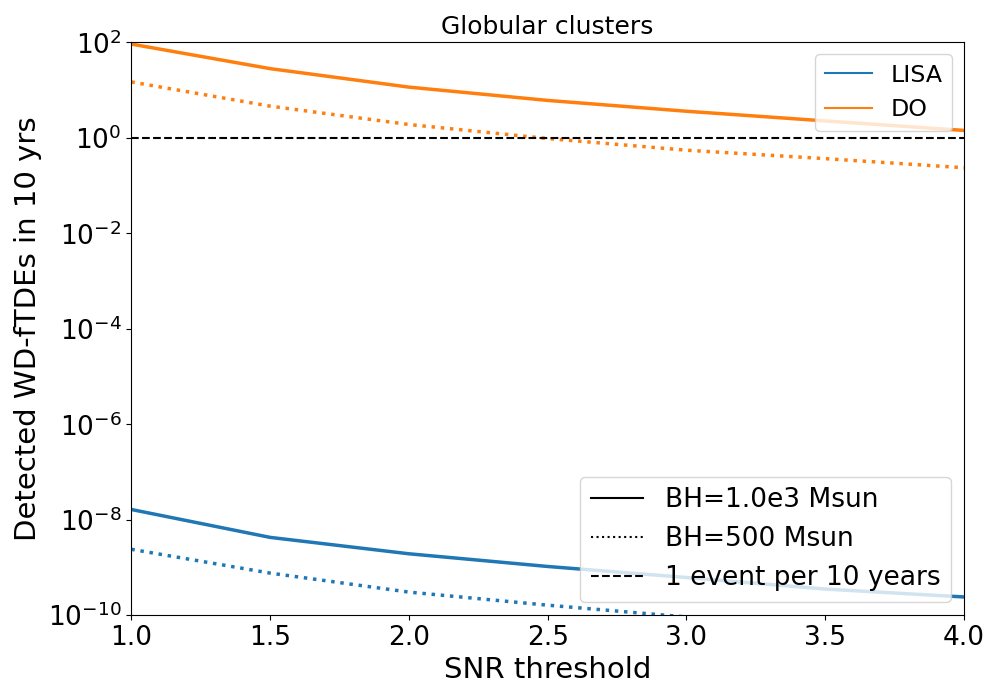}
    \end{minipage}
    \hfill
    \begin{minipage}{0.49\textwidth}
        \centering
        \includegraphics[width=\linewidth]{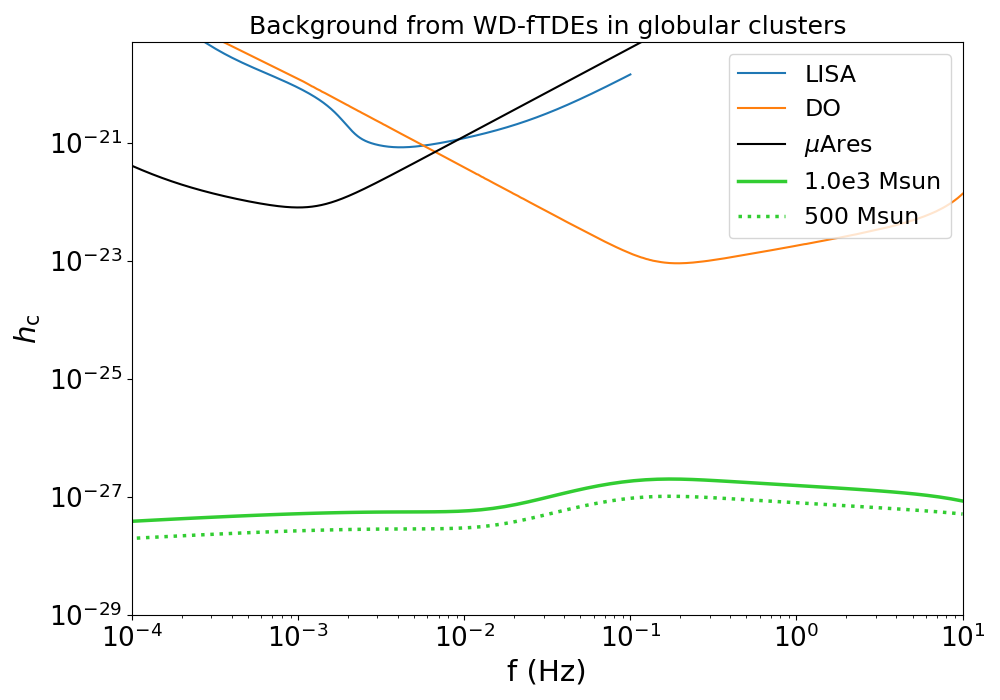}
    \end{minipage}
    \caption{\textcolor{black}{Left panel: detection rates of WD-fTDEs from globular clusters plotted for different SNR thresholds. Right panel: GWB from WD-fTDEs from globular clusters. The color scheme and the detectors are the same as in Fig. \ref{fig:wd_individual_det}-\ref{fig:wd_back}. The solid line represents an IMBH of $10^3\text{M}_{\odot}$, while the dotted line an IMBH of $500\text{M}_{\odot}$.}}
    \label{fig:500_WDFTDEs}
\end{figure*}
\textcolor{black}{We have also investigated the GW signal from a population of WD–fTDEs occurring in GCs hosting black holes with masses of 
$500 \text{M}_{\odot}$. As shown in Fig. \ref{fig:500_WDFTDEs}, both the detection rates and the background are generally lower than those presented earlier in the main text, making the detection of such events even less likely. However, obtaining a more realistic estimate will ultimately require observational constraints on the IMBH population in GCs.}
\end{appendix}
\end{document}